\documentclass[11pt]{article}

\usepackage{amsmath}
\usepackage{amssymb}
\usepackage{graphicx,color}
\usepackage{wrapfig,framed}
\usepackage{fixltx2e}
\usepackage[titletoc, title]{appendix}
\numberwithin{equation}{section}

\usepackage{color}

\newcommand{\overbar}[1]{\overline{\mkern-4mu#1\mkern-0.0mu}}

\newcommand{\p}{\partial}
\newcommand{\pal}{\partial}
\newcommand{\ve}{\epsilon}
\newcommand{\e}{\epsilon}
\newcommand{\nn}{\nonumber}

\newcommand{\F}{\mathcal{F}}

\newcommand{\V}{\mathcal{V}}
\newcommand{\PP}{\mathcal{P}}
\renewcommand{\H}{\mathcal{H}}

\newcommand{\lm}{\Lambda}
\newcommand{\bt}{{\bf t}}
\newcommand{\tr}{{\rm tr}}

\newcommand{\Zgue}{Z_{\textsc{\tiny\rm GUE}}}
\newcommand{\Fgue}{\mathcal{F}_{\textsc{\tiny\rm GUE}}}

\newcommand{\eqa}{\begin{eqnarray}}
\newcommand{\eeqa}{\end{eqnarray}}
\newcommand{\beq}{\begin{equation}}
\newcommand{\eeq}{\end{equation}}

\newtheorem{theorem}{Theorem}[section]
\newtheorem{cnj}[theorem]{Conjecture}
\newtheorem{rmk}[theorem]{Remark}
\newtheorem{prp}[theorem]{Proposition}
\newtheorem{lem}[theorem]{Lemma}
\newtheorem{cor}[theorem]{Corollary}

\newtheorem{emp}[theorem]{Example}

\newenvironment{prf}{\noindent {\it Proof.} }{\hfill $\Box$}

\begin{document}

\title{Hodge--GUE Correspondence and the Discrete KdV Equation}
\author{
{Boris Dubrovin$^{*}$, Si-Qi Liu$^{**}$, Di Yang$^\dagger$, Youjin Zhang$^{**}$}\\
{\small ${}^{*}$ SISSA, via Bonomea 265, Trieste 34136, Italy}\\
{\small ${}^{**}$ Department of Mathematical Sciences, Tsinghua University, Beijing 100084, P.R.~China}\\
{\small ${}^\dagger$ School of Mathematical Sciences, University of Science and Technology of China,} \\
{\small Hefei 230026, P.R.~China}}
\date{}\maketitle

\begin{abstract} 
We prove the conjectural relationship recently proposed in~\cite{DY2} between certain special cubic 
Hodge integrals of the Gopakumar--Mari\~no--Vafa type \cite{GV, MV} 
and GUE correlators, and the conjecture
proposed in~\cite{DLYZ} that the partition function of these Hodge 
integrals is a tau function of the discrete KdV hierarchy.
\end{abstract}
{\small \noindent{\bf Mathematics Subject Classification (2010).} 53D45; 37K10, 15B52.}

\tableofcontents

\section{Introduction}\label{section1}
Let ~$\overbar{\mathcal{M}}_{g, k}$ denote the 
Deligne--Mumford moduli space of stable algebraic curves of genus $g$ with $k$ distinct marked points. Denote by $\mathcal{L}_i$ the $i^{th}$ tautological line bundle over ~$\overbar{\mathcal{M}}_{g, k}$, 
and by $\mathbb{E}_{g, k}$ the rank $g$
Hodge bundle. Denote $\psi_i:=c_1(\mathcal{L}_i)$, $i=1,\dots, k$, and $\lambda_j:= c_j(\mathbb{E}_{g, k})$, $j=0, \dots, g$. 
The Hodge integrals are integrals over ~$\overbar{\mathcal{M}}_{g, k}$ of the form
\[
\int_{\overbar{\mathcal{M}}_{g,k}} \psi_1^{i_1}\cdots \psi_k^{i_k} \lambda_1^{j_1} \cdots \lambda_g^{j_g}, \quad i_1, \dots, i_k,\, j_1, \dots, j_g \geq 0. 
\]
These integrals will be defined to be {\em zero} unless
\[
3g-3+k = \sum_{\ell=1}^k i_\ell + \sum_{\ell=1}^g \ell \, j_\ell. 
\]
We consider in this paper the following {\em special cubic Hodge integrals}
\beq\label{chbcy}
\int_{\overbar{\mathcal{M}}_{g,k}} \Lambda_g(p) \Lambda_g(q) \Lambda_g(r)\psi_1^{i_1}\cdots \psi_k^{i_k}, 
\eeq
where $\Lambda_g(z):= \sum_{j=0}^g \lambda_j  z^j$ denotes the Chern polynomial of $\mathbb{E}_{g,k}$, 
and $p, q, r\in\mathbb{C}$ satisfy the {\em local Calabi--Yau condition}
\begin{equation}\label{zh-23}
\frac{1}{p}+\frac{1}{q}+\frac{1}{r}=0 .
\end{equation}
The special cubic Hodge integrals play important roles in the localization computation of the Gromov--Witten
invariants of toric Calabi--Yau threefolds \cite{GP, LLZ, LLZ2}, the theory of topological vertex \cite{AKMV, LLLZ}, 
and the BKMP remodeling conjecture \cite{BKMP1, BKMP2, EO, FLZ}.  We mainly consider the case with
$p=q$ in the present paper, and leave the general case to a subsequent publication. 
Note that the cubic Hodge integrals satisfy the following homogeneity condition:
\begin{align*}
& \int_{\overbar{\mathcal{M}}_{g, k}} \Lambda_g(\rho p) \Lambda_g(\rho q)\Lambda_g(\rho r) \psi_1^{i_1}\cdots \psi_k^{i_k} \\
&= \rho^{3g-3+k-i_1-\cdots-i_k}\int_{\overbar{\mathcal{M}}_{g, k}} \Lambda_g(p)\Lambda_g(q) \Lambda_g(r) \psi_1^{i_1}\cdots \psi_k^{i_k},
\end{align*}
so we only need to consider the following special Hodge integrals:
\begin{equation}\label{zh-22}
\int_{\overbar{\mathcal{M}}_{g, k}} \Lambda_g(-1) \Lambda_g(-1) \Lambda_g\bigl(\tfrac{1}{2}\bigr) \psi_1^{i_1}\cdots \psi_k^{i_k}.
\end{equation}
We call the following generating function the {\it special cubic Hodge free energy}:
\begin{equation} \label{cubic-hodge}
\mathcal H_{\tiny\rm cubic}(\bt;\epsilon) = \sum_{g\geq 0} \epsilon^{2g-2} \sum_{k\geq 0}  \frac 1{k!} 
\sum_{i_1, \dots, i_k\geq 0} t_{i_1}\cdots t_{i_k}\int_{\overbar{\mathcal M}_{g, k}} 
\Lambda_g(-1) \Lambda_g(-1) \Lambda_g\bigl(\tfrac12\bigr) \psi_1^{i_1}\cdots \psi_k^{i_k}.
\end{equation}
Here $\bt=(t_0, t_1, \dots)$ are independent variables, and $\epsilon$ is a parameter often called the string coupling constant. 
Denote by $\mathcal{H}_g=\H_g(\bt)$ the genus $g$ part of 
$\mathcal H_{\tiny\rm cubic}(\bt;\epsilon)$ so that
\beq\label{genusexpand}
\mathcal{H}_{\tiny\rm cubic}(\bt;\e) = \sum_{g\geq0} \epsilon^{2g-2} \mathcal{H}_g(\bt).
\eeq
We also call the exponential 
\beq
e^{\mathcal H_{\tiny\rm cubic}(\bt;\e)}=: Z_{\tiny\rm cubic}(\bt;\e)
\eeq
the special cubic Hodge partition function. 

On the other hand, let ${\mathcal H}(N)$ be the space of $N\times N$ Hermitian matrices. Denote by
\[
dM = \prod_{i=1}^N d M_{ii} \prod_{i<j} d{\rm Re} M_{ij} d{\rm Im}M_{ij}
\]
the standard unitary invariant volume element on  ${\mathcal H}(N)$. The GUE partition function of size $N$ with even couplings is defined by
\beq\label{part}
Z_N({\bf s};\e) = \frac{(2\pi)^{-{N}} }{{{\rm Vol}}(N)} \int_{{\mathcal H}(N)} e^{- {\frac1\e}  \tr V(M;\, {\bf s})} dM.
\eeq
Here $V(M;  {\bf s})$ is an even polynomial, or more generally a power series in $M$ of the form
\beq\label{pot}
V(M; {\bf s}) = \frac12 M^2 -\sum_{j\geq 1} s_{2j}  M^{2j}
\eeq
with ${\bf s}:=(s_2, s_4, s_6, \dots)$,  
and $\textrm{Vol}(N)$ is the volume of the quotient of the unitary group 
over the maximal torus $\bigl[ U(1)\bigr]^N$, which is given by
$$
\textrm{Vol}(N) = \textrm{Vol}\left( U(N)/\left[ U(1)\right]^N\right) = \frac{\pi^{\frac{N(N-1)}2} }{G(N+1)}, \quad G(N+1):=\, \prod_{n=1}^{N-1} n! .
$$

The integral that appears in \eqref{part} is considered as a formal power series of $s_2, s_4, s_6, \dots$,
whose coefficients are analytic in $N$.  
Introduce the  {\em 't Hooft coupling parameter} $x$ by
\begin{equation}\label{zh-36}
x :=N \epsilon.
\end{equation}
Expanding the free energy $\mathcal{F}_N({\bf s};\e):=\log Z_N({\bf s};\e)$ in powers of $\epsilon$ and 
replacing the Barnes $G$-function $G(N+1)$ by its asymptotic expansion in $\e$ yields
\beq\label{genusF}
\mathcal{F}_{\tiny\rm even}(x, {\bf s}; \epsilon) 
:=\mathcal{F}_N({\bf s}) |_{N=\frac{x}{\epsilon}} -\frac1{12}\log \epsilon = \sum_{g\geq 0} \epsilon^{2g-2} \mathcal{F}_g (x, {\bf s}).
\eeq
The GUE free energy $\mathcal{F}_{\tiny\rm even}(x, {\bf s}; \epsilon)$ with even couplings can be represented \cite{BIZ, HZ, thooft1, thooft2, mehta} in the form
\begin{align}
&\mathcal{F}_{\tiny\rm even}(x, {\bf s};\epsilon ) =  \frac{x^2}{2\epsilon^2} \left( \log x -\frac32\right) -\frac1{12} \log x +\zeta'(-1) +\sum_{g\geq 2} \epsilon^{2g-2} \frac{B_{2g}}{4g(g-1)x^{2g-2}} \nn\\
 & \qquad\qquad\qquad\quad + \sum_{g\geq 0} \epsilon^{2g-2} \sum_{k\geq 0} \sum_{i_1, \dots, i_k\geq 1} a_g(2i_1, \dots, 2i_k)  s_{2i_1} \dots s_{2i_k} x^{2-2g -(k-{|i|})}, \label{expan}
\end{align}
where
$$
a_g(2i_1, \dots, 2i_k) = \sum_{\Gamma} \frac{1}{\#\, {\rm Sym} \Gamma}
$$
and the last summation is taken over all connected oriented ribbon graphs $\Gamma$ 
(with labelled half edges and unlabelled vertices) of genus $g$ with $k$ vertices of valencies $2i_1$, \dots, $2i_k$,  $|i|:=i_1+\dots + i_k$, and
$\#\,{\rm Sym} \Gamma$ is the order of the symmetry group of $\Gamma$. Here and in what follows, $B_k$ are the Bernoulli numbers.
The exponential of the GUE free energy
\beq \label{GUE-even-Z}
e^{\F_{\tiny\rm even}(x,{\bf s};\e)} =: Z_{\tiny\rm even}(x,{\bf s};\e)
\eeq
is called the GUE partition function with even couplings. It is convenient to change normalization of the even couplings by introducing
\beq\label{ssbar}
\bar{s}_{k}:=\binom{2k}{k} s_{2k},\quad k\ge 1.
\eeq

The following statement was formulated in~\cite{DY2}.

\begin{cnj}  \label{conjecture1} 
The GUE partition function with even couplings and the special Hodge partition function are related by the identity
\beq\label{id1}
Z_{\tiny\rm even}(x, {\bf s}; \epsilon) = e^{\frac{A(x, {\bf s})}{\epsilon^2}+\zeta'(-1)}  
Z_{\tiny\rm cubic}\left({\bf t}\left( x+\tfrac{\e}2, {\bf s}\right); \sqrt{2} \epsilon\right)  Z_{\tiny\rm cubic}\left({\bf t}\left( x-\tfrac{\e}2, {\bf s}\right); \sqrt{2} \epsilon\right),
\eeq
where
\beq\label{axs}
A(x, {\bf s}) =
\frac12  \sum_{k_1, k_2\ge 1} \frac{k_1 k_2}{k_1+k_2} \bar{s}_{k_1} \bar{s}_{k_2} - \sum_{k\ge 1} \frac{k}{1+k} \bar{s}_{k} +
x  \sum_{k\ge 1} \bar{s}_{k}   + \frac14  - x
\eeq
and 
\beq\label{time-sub}
t_i(x, {\bf s}) :=\, \sum_{k\geq 1} k^{i+1} \bar{s}_k  - 1 + \delta_{i,1} + x  \delta_{i,0},\quad i\geq 0.
\eeq
\end{cnj}

\begin{rmk}
The left hand side of the logarithm of~\eqref{id1} is a formal power series in~$x-1$ 
and $\bar{s}_1, \bar{s}_2, \dots$. If one
expands the right hand side as formal power series in $x-1$ and $\bar{s}_1, \bar{s}_2, \dots$, the coefficients
are infinite sums of Hodge integrals. For example, the constant term of the logarithm of~\eqref{id1}
is a formal Laurent series of~$\e$ with powers greater or equal to~$-2$. Acting by the operator
\[\frac{1}{e^{\e \p_x/2}+e^{-\e \p_x/2}}\]
on this constant term, and then taking the coefficient of $\e^{2g-2}\ (g\ge 2)$), we obtain
\begin{align}
&\frac1{2g(2g-1)(2g-2)} \sum_{g'=0}^g  (2g'-1) 
\binom{2g}{2g'}
\frac{E_{2g-2g'} B_{2g'}} {2^{2g-2g'}}\nn\\
&= 2^g \sum_{ \mu \in \mathbb{Y}}   \frac{(-1)^{\ell(\mu)}}{m(\mu)!}  \int_{\overbar{\mathcal M}_{g,\ell(\mu)}} 
\Lambda_g(-1) \Lambda_g(-1) \Lambda_g \bigl(\tfrac12\bigr)
\prod_{i=1}^{\ell(\mu)} \psi_i^{\mu_i+1}.   \label{simpleid}
\end{align}
Here $\mathbb{Y}$ denotes the set of partitions of non-negative integers, $\ell(\mu)$ denotes the length
of $\mu \in \mathbb{Y}$, $m_i(\mu)$ denotes the multiplicity of $i$ in $\mu$,
 $m(\mu)!:=\prod_{i=1}^\infty m_i(\mu)!$, and $E_k$ are the Euler numbers defined by the generating function
$$
 \frac{2}{e^z+e^{-z}} = \sum_{k=0}^\infty \frac{E_k} {k!} z^k.
$$
Note that the right hand side of~\eqref{simpleid} is actually a finite sum due to the dimension reason,
so these terms are well-defined. Due to similar reasons, each Taylor coefficient of the right hand side of the logarithm of~\eqref{id1} is
well-defined.
\end{rmk}

We refer to the conjectural identity~\eqref{id1} as a {\em Hodge--GUE correspondence}. 

The Hodge--GUE correspondence conjecture was verified in~\cite{DY2} for $g=0,1,2$. 
In the present paper we prove it for any genus.

\begin{theorem} [Main Theorem]  \label{thm1}
The Conjecture \ref{conjecture1} holds true.
\end{theorem}

The proof will be given in Sections \ref{section2}, \ref{virgue}, \ref{section4} below. 
For the readers' convenience, 
we outline here the sketch and the main ideas of our proof. The first step is to show that both sides of \eqref{id1} 
after suitable and explicit transformations satisfy {\it the same} 
linear constraints defined by explicit linear operators, which will be called the Virasoro constraints. The next step is to prove a certain {\it uniqueness theorem} for solutions of the Virasoro constraints. 
Our idea of proving the uniqueness theorem is as follows: the higher genus terms in the genus expansions of the logarithms of both sides of \eqref{id1} admit the so-called jet-variable representations. 
The same linear constraints then lead to the same {\em loop equation} for the higher genus terms as functions of the jet variables. 
The uniqueness theorem is then established by the uniqueness of solution of the loop equation, which completes the proof. 
We note that this idea was used in the study of the so-called Gromov--Witten potentials \cite{DZ-norm}. The loop equation for the special cubic Hodge potentials, which is given in \eqref{Hgeq2}, can also be used to compute the Hodge integrals as we show in the following proposition.
\begin{prp} \label{loophodgeprop}
Let us denote
\beq\label{v-dispkdv-intro}
v(\bt) =\sum_{k\ge 1}\frac{1}{k} \sum_{i_1+\cdots+i_k=k-1}\frac{t_{i_1}}{i_1!}\cdots\frac{t_{i_k}}{i_k!},
\eeq
and $v_k=\p_{t_0}^k v({\bf t}), \, k\ge 0$. Then there is a unique sequence of functions $\{H_g(v,v_1,\dots,v_{3g-2})\}_{g\ge1}$
satisfying the following recursion equations:
\begin{align}
& H_1=\frac{1}{24}\log v_1-\frac{1}{16}v, \\
& \sum_{j=1}^{3g-2} j  v_j \frac{\p H_g}{\p v_j} = (2g-2) \, H_g, \quad g\geq 2, \label{Hgeq1} \\
& \sum_{q\geq 0} 
\left(\p_{t_0}^q \left(\frac1{B^2}\right) + \sum_{r=1}^q \binom{q}{r} \p_{t_0}^{r-1} \left(\frac1B\right) \p_{t_0}^{q-r+1} \left(\frac1B\right)\right)  
\frac{\p \triangle H}{\p v_q} \nn\\
& \quad -\frac{\e^2}2 \sum_{q_1, q_2\ge 0} \p_{t_0}^{q_1+1} 
\left(\frac1B\right) \p_{t_0}^{q_2+1} \left(\frac1B\right)  \left(\frac{\p \triangle H}{\p v_{q_1}}  \frac{\p \triangle H}{\p v_{q_2}}  + \frac{\p^2 \triangle H}{\p v_{q_1} \p v_{q_2}}\right)  \nn\\
& \quad -\frac{\e^2}2 \sum_{q\ge 0} \p_{t_0}^{q+2} \left(\frac1{8B^4} - \frac1{4B^2} \right) \, \frac{\p \triangle H }{\p v_q} 
+ \frac1 {8B^2} - \frac1{16B^4} = 0  \label{Hgeq2}
\end{align}
with $\triangle H := \sum_{g\geq1} \e^{2g-2} H_g$,
$B= \sqrt{1-\frac{4e^v}{\lambda}}$. Moreover, the special cubic Hodge potentials can
be represented as
$$ \H_g(\bt) = H_g\Biggl( v(\bt ), \frac{\pal v(\bt )}{\pal t_0}, \dots, \frac{\pal^{3g-2} v(\bt)}{\pal t_0^{3g-2}}\Biggr) , 
\quad g\geq 1.  $$
\end{prp}

This proposition follows from Propositions \ref{virasorowfstarthm}, \ref{zh-11}, \ref{zh-12} of Section~\ref{section4}. From the Main Theorem, we also have the following corollary which relates the special cubic Hodge integrals with the discrete KdV hierarchy.
\begin{cor}\label{zh-15}
Introduce a shift operator $\lm_1=e^{\e\p_{t_0}}$, and define
\begin{align}
&W({\bf t}; \e)=\biggl(\lm_1^{\frac{1}{2}}+\lm_1^{-\frac{1}{2}}\biggr)\Bigl(\lm_1-2+\lm_1^{-1}\Bigr)\H_{\tiny\rm cubic}(\bt;\sqrt{2} \e),\label{zh-16}\\
&\frac{\p}{\p T_k}=\sum_{i\ge0}k^{i+1}\frac{\p}{\p t_i},\quad k\ge 1.\label{zh-17}
\end{align}
Then we have
\begin{align*}
2\e \frac{\p W}{\p T_1}=&\left(\lm_1-\lm_1^{-1}\right)\exp{(W)},\\
6 \e\frac{\p W}{\p T_2}=& \left(\lm_1-\lm_1^{-1}\right)\left(\exp(W)\left(\lm_1+1+\lm_1^{-1}\right)\exp(W)\right),\dots.
\end{align*}
More generally, $\frac{\p W}{\p T_k}$ is given by the $k$-th equation of the discrete KdV hierarchy (also called the Volterra lattice hierarchy)
\begin{equation}\label{zh-21}
\binom{2k}{k} \e \, \frac{\p L_1}{\p T_k}=\biggl[\left(L_1^{2k}\right)_+,\ L_1\biggr],
\end{equation}
where $L_1=\lm_1+\exp(W)\lm_1^{-1}$.
\end{cor}

The paper is organized as follows: In Sections \ref{section2}, \ref{virgue}, \ref{section4} we give the proof of the Main Theorem. 
In Section \ref{zh-25} we give the proof of Corollary \ref{zh-15}. In Section \ref{zh-26} we give a brief review of an application of the Main Theorem. 
In the Appendix we present some Givental quantization formulae that are used in this paper. 

\paragraph{Acknowledgements.} We are grateful to Don Zagier and Jian Zhou for several very useful discussions. 
This work is partially supported by NSFC No.\,11771238, No.\,11725104 and No.\,11371214. The work of B.D. is partially supported 
by the Russian Science Foundation Grant No. 16-11-10260 ``Geometry and Mathematical Physics of Integrable Systems". 
Parts of the work of D.Y. were done in SISSA, Trieste and in MPIM, Bonn while he was a postdoc; 
he acknowledges both SISSA and MPIM for excellent working conditions and supports.

\section{Virasoro constraints for the cubic Hodge partition function} \label{section2}
By using the results of~\cite{FP}, it is shown in~\cite{DLYZ}
that the special cubic Hodge partition function has the following expression:
\beq\label{zh-28}
Z_{\tiny\rm cubic} (\bt; \e)= \exp \left(\sum_{j = 1}^\infty  \frac{B_{2j}}{j (2j-1)}  (2^{-2j}-1) D_j(\e^{-1} \tilde \bt, \e \p/\p \bt) \right) Z(\bt; \e). 
\eeq
Here 
\beq\label{dj}
D_j(\e^{-1}\bt, \e \p/\p \bt):= - \sum_{i\geq 0} t_i \frac{\p }{\p t_{i+2j-1}} + \frac{\e^2}2 \sum_{a=0}^{2j-2} (-1)^a \frac{\p^2 }{\p t_a \p t_{2j-2-a}},\quad j\geq 1,
\eeq
$\tilde{\bf t}=\bigl(\tilde{t}_0, \tilde{t}_1,\dots\bigr)$ is defined by
the dilaton shift
\begin{equation}\label{zh-1}
\tilde t_i=t_i-\delta_{i,1},\quad i\ge 0,
\end{equation} 
and $Z$ is the partition function of 2d topological gravity~\cite{Witten} given by 
$$
Z= Z(\bt;\e) := \exp\left(\sum_{g\geq 0} \e^{2g-2} \sum_{k\geq 0} \frac1{k!} \sum_{i_1, \dots, i_k\geq 0} \int_{\,\overbar{\mathcal{M}}_{g, k}} \psi_1^{i_1}\cdots \psi_k^{i_k}  t_{i_1}\cdots t_{i_k}\right).
$$
It is well known that $Z$ satisfies the following Virasoro constraints \cite{DVV, Kontsevich, Witten}:
\beq \label{Vira-WK}
L_m\left( \e^{-1}\tilde{\bf t}, \e \p/\p{\bf t}\right) Z({\bf t}; \epsilon) = 0, \quad m\geq -1,
\eeq
where the Virasoro operators $L_m=L_m\left( \e^{-1}{\bf t}, \e \p/\p{\bf t}\right)$, $m\geq -1$ are given by
\begin{align}
& L_{-1} = \sum_{i\geq 1} t_i \frac{\pal}{\pal t_{i-1}} +\frac{t_0^2}{2 \epsilon^2}, \label{virakdvintro1} \\
& L_0 =\sum_{i\geq 0} \frac{2i+1}{2} t_i \frac{\pal}{\pal t_{i}}+\frac{1}{16},  \label{virakdvintro2}\\
& L_m =\frac{\epsilon^2}{2} \sum_{i+j=m-1} \frac{(2i+1)!! (2j+1)!!}{2^{m+1}}
\frac{\pal^2}{\pal t_i \pal t_j}+ \sum_{i\geq 0} \frac{(2i+2m+1)!!}{2^{m+1}
(2i-1)!!} t_i \frac{\pal}{\pal t_{i+m}}, 
\quad m\geq 1. \label{virakdvintro3}
\end{align}
They satisfy the following Virasoro commutation relations:
\beq
\left[L_m, L_n \right] = (m-n) L_{m+n}, \quad \forall\, m, n\geq -1. \label{viracommkdv}
\eeq

Define the operators $L_m^{\rm cubic}$ by 
\begin{equation}
L_m^{\rm cubic}\left( \e^{-1}{\bf t}, \e \p/\p{\bf t}\right) = e^{G}  \circ L_m\left(\e^{-1}{\bf t}, \e \p/\p{\bf t}\right) \circ e^{- G}, \quad m\ge -1,\label{deflmcubic}
\end{equation}
where
\[G:=\sum_{j = 1}^\infty \frac{B_{2j}}{j (2j-1)} \left(2^{-2j}-1\right)  D_j, \quad m\geq -1.\]
By a straightforward calculation we obtain that \cite{DLYZ, Zhou1}
\beq\label{zh-30}
L_{-1}^{\rm cubic} = \sum_{k\geq 1} t_k \frac{\p}{\p t_{k-1}}+\frac{ t_0^2}{2\ve^2}-\frac1{16} .
\eeq
It follows from \eqref{viracommkdv} that
$$
\left[L_m^{\rm cubic},L_n^{\rm cubic}\right] = (m-n) \, L_{m+n}^{\rm cubic},\quad \forall\,m, n\geq -1.
$$
By using \eqref{zh-28} and \eqref{Vira-WK} we also obain the following lemma.
\begin{lem} [\cite{Zhou1}] \label{vira-v1}
The special cubic Hodge partition function satisfies the Virasoro constraints
\beq\label{vira-Hodge}
L_m^{\rm cubic} \left( \e^{-1}\tilde{\bf t}, \e \p/\p{\bf t}\right) Z_{\tiny\rm cubic}({\bf t};\e) = 0,\quad \forall\, m\geq -1.
\eeq
\end{lem}

Let us proceed to give a second version of Virasoro constraints for the special cubic Hodge partition function $Z_{\tiny\rm cubic}({\bf t};\e)$ which is crucial to establish, 
in Section~\ref{section4},  its relation with the GUE partition function. These Virasoro
constraints are obtained by linear combinations of the ones given in~\eqref{vira-Hodge}, and the associated Virasoro operators are defined as follows:
\begin{equation}\label{wlmdef}
\widetilde L^{\rm cubic}_{m}\left( \e^{-1} {\bf t}, \e \p/\p{\bf t}\right) :=\sum_{k=-1}^\infty \frac{m^{k+1}}{(k+1)!} L_k^{\rm cubic}\left( \e^{-1} {\bf t}, \e \p/\p{\bf t}\right),\quad m\geq 0 . 
\end{equation}
In order to obtain the commutation relations of the operators  $\widetilde L^{\rm cubic}_{m}$, we need to use the following lemma which can be proved by a straightforward calculation.
\begin{lem} \label{alg_lem}
Let $\{a_m \, | \, m\geq -1\}$ be a basis of an infinite dimensional Lie algebra satisfying
the commutation relations
$$
[a_m, a_n] = (m-n) a_{m+n} \quad \forall \, m, n\geq -1,
$$ 
where $[\ ,\,]$ denotes the Lie bracket of the Lie algebra. Define 
$$
\tilde a_{m} := \sum_{k\geq -1} \frac{m^{k+1}}{(k+1)!}\,  a_k,\quad m\geq 0. 
$$
Then
$$
\left[ \tilde a_{m}, \tilde a_{n}\right] = (m-n) \tilde a_{m+n}, \quad \forall \, m, n\geq 0 . 
$$
\end{lem}

By using the above lemma and Lemma \ref{vira-v1}, we obtain the following corollary.

\begin{cor}\label{vira-2-lem}
The special cubic Hodge partition function satisfies the following constraints:
\beq\label{vira-Hodge2}
\widetilde L_m^{\rm cubic} \bigl( \e^{-1}\tilde{\bf t}, \e \p/\p{\bf t}\bigr) Z_{\tiny\rm cubic} = 0, \quad \forall\, m\geq 0.
\eeq
Moreover, the operators $\widetilde L_m^{\rm cubic}$ satisfy the Virasoro commutation relations
\begin{equation}\label{zh-6}
\left[ \widetilde L_{m}^{\rm cubic}, \widetilde L_{n}^{\rm cubic}\right] = (m-n) \widetilde L_{m+n}^{\rm cubic}, \quad \forall \, m, n\geq 0 . 
\end{equation}
\end{cor}

We call \eqref{vira-Hodge2} the 2nd version of Virasoro constraints for~$Z_{\tiny\rm cubic}$.

\begin{theorem}  \label{lwl0-wl1-wl2} 
We have the following explicit expressions of the first three Virasoro operators:
\begin{align}
&\widetilde L_0^{\rm cubic}\left( \e^{-1}{\bf t}, \e \p/\p{\bf t}\right) =  \sum_{i\ge 1} t_i \frac{\p }{\p t_{i-1}}+\frac{t_0^2}{2\ve^2} -\frac1{16},  \label{wl0}\\
&\widetilde L_1^{\rm cubic}\left( \e^{-1}{\bf t}, \e \p/\p{\bf t}\right) =  \frac12\sum_{i\geq 0} \sum_{j=0}^i  \binom{i}{j} \left(2 t_{j+1}+  t_j  \right)\frac{\p }{\p t_i} + \frac{t_0^2}{2\e^2},\label{wl1}\\
& \widetilde L_2^{\rm cubic}\left( \e^{-1}{\bf t}, \e \p/\p{\bf t}\right) =  \frac{\e^2}8 \sum_{i, j\geq 0} \frac{\p^2 }{\p t_i \p t_j} + \sum_{i\geq 0} \sum_{j=0}^i \binom{i}{j}2^{i-j}  \left({t}_{j+1}+{t}_{j}\right) \frac{\p}{\p t_i} \nn\\
& \quad - \frac18\sum_{i\geq 1} \sum_{j=0}^{i-1} \sum_{r=0}^{i-1-j} (-1)^r 
\binom{i}{i-1-j-r}2^{i-j-r} t_j \frac{\p }{\p t_i}+\frac{ t_0^2}{2\e^2} + \frac1{16}. \label{wl2}
\end{align}
\end{theorem}
\begin{prf} From the definition \eqref{wlmdef} it follows that $\widetilde L_0^{\rm cubic}=L_{-1}^{\rm cubic}$. So by using \eqref{zh-30} we get the formula \eqref{wl0}.

For $m>0$, the direct calculation of~$\widetilde L_m^{\rm cubic}$ becomes complicated, so we are to use Givental's quantization formulae (see Appendix \ref{appa}) to simplify the computations. To this end, let us introduce a function 
\beq\label{Gamma}
\Phi(z) = 2^{-2z}\, \frac{\Gamma(1-z)}{\Gamma(1+z)}\sqrt{\frac{\Gamma(1+2z)}{\Gamma(1-2z)}}\ .
\eeq
It is a multivalued meromorphic function of $z\in\mathbb C$ with branch points at nonzero half-integers. With a suitable choice of the branches one has
$$
\Phi(z) \to e^{\mp \frac{\pi i}4}, \quad\lvert z \rvert \to \infty,\, 
\pm \operatorname{Re}(z)>0.
$$
It satisfies the identity
$$
\Phi(-z) \Phi(z) = 1,
$$
and so it defines a canonical transformation
$$
f(z) \, \mapsto \, \Phi(z) f(z)
$$
on the Givental symplectic space (see Appendix~\ref{appa} for the details about Givental quantization formalism). 
We now identify the function $\Phi(z)$ with its asymptotic expansion at $\lvert z \rvert \to \infty$ (see Appendix~\ref{appa}).
Denote by $\widehat\Phi$ the quantization of this symplectomorphism acting on the corresponding Fock space. By using Lemma~\ref{g3} we obtain
$$
e^G = \widehat\Phi.
$$
So the operators~$L_k^{\rm cubic}$ defined in~\eqref{deflmcubic} have the expressions 
$$
L_k^{\rm cubic} \left( \e^{-1}{\bf t}, \e \p/\p{\bf t}\right) = \widehat\Phi L_k \left( \e^{-1}{\bf t}, \e \p/\p{\bf t}\right) \widehat{\Phi}^{-1}, \quad k\geq -1.
$$
Thus by using Lemma~\ref{g1} we obtain
\beq \label{tobesim}
L_k^{\rm cubic} \left(\e^{-1}{\bf t}, \e \p/\p{\bf t}\right)= 
\left. \left[\widehat{\Phi}\left(\widehat{l_k} + \frac{\delta_{k, 0}}{16}\right) \widehat{\Phi}^{-1} \right]\right|_{q_i\mapsto  t_i, \, \p_{q_i} \mapsto \p_{t_i},\,i\geq 0} , \quad k\geq -1,
\eeq
where
\begin{equation}\label{lklk}
l_k= (-1)^{k+1}  z^{3/2} \p_z^{k+1} z^{-1/2}, \quad k\geq -1.
\end{equation}
Denote $\Phi_k(z)=\Phi(z) l_k \Phi(z)^{-1}$, then from~\eqref{tobesim} we have
$$ 
L_k^{\rm cubic} \left( \e^{-1}{\bf t}, \e \p/\p{\bf t}\right) =  
\widehat{\Phi}_k|_{q_i\mapsto t_i, \, \p_{q_i} \mapsto \p_{t_i},\,i\geq 0} + \frac{\delta_{k,0}}{16} - \frac{ \delta_{k, -1} }{16},\quad k\geq -1 .
$$
Here we have used the fact that the non-zero cocycle terms of the above quantization formula
appear only when~$k=-1$.  
Thus the operators $\widetilde L_m^{\rm cubic},\,m\geq 0$ have the following expressions:
\beq \label{wlm}
\widetilde L_m^{\rm cubic}\left( \e^{-1}{\bf t}, \e \p/\p{\bf t}\right) = 
\widehat{\Psi}_m|_{q_i\mapsto t_i, \, \p_{q_i} \mapsto \p_{t_i},\,i\geq 0}  + \frac{m-1}{16},
\eeq
where 
\[\Psi_m(z)=\sum_{k=-1}^\infty \frac{m^{k+1}}{(k+1)!}\Phi_k(z).\]

Now let us employ the quantization formulae~\eqref{wlm} of the operators $\widetilde{L}_m^{\rm cubic}$ to prove the theorem.
From~\eqref{lklk} we have
$$
\Psi_1(z)=  \Phi(z) z^{3/2} e^{-\p_z} z^{-1/2}\Phi^{-1}(z) = z^{3/2} (z-1)^{-1/2} \frac{\Phi(z)}{\Phi(z-1)} e^{-\p_z}.
$$
By using the identity
\beq\label{ratio}
\frac{\Phi(z)}{\Phi(z-1)} = \frac{z-\frac12}{\sqrt{z(z-1)}},
\eeq
we arrive at
$$\Psi_1(z)= z \frac{z-1/2} {z-1}  e^{-\p_z} .  
$$
In order to prove the formula~\eqref{wl1}, we need to compute the residue
$$
-\frac12 \operatorname{Res}_{z=\infty }f(-z) \frac{z-1/2} {z-1}f(z-1) \frac{dz}{z}
$$
for $f(z)=q(z)+p(z)$, where $q(z)= \sum_{i\geq 0} q_i \, z^{-i}$ and $p(z)= \sum_{i\geq 0} p_i \, (-z)^{i+1}$. 
The computation of the above residue can be decomposed into the following four parts: 
\begin{align}
&S_1= -\frac12 \operatorname{Res}_{z=\infty} p(-z)\frac{z-1/2} {z-1}p(z-1)\frac{dz}{z},\quad
S_2 = -\frac12 \operatorname{Res}_{z=\infty } p(-z)\frac{z-1/2} {z-1}q(z-1)\frac{dz}{z}, \nn\\
&S_3 =-\frac12 \operatorname{Res}_{z=\infty }q(-z)\frac{z-1/2} {z-1} p(z-1) \frac{dz}{z},\quad 
S_4 =-\frac12 \operatorname{Res}_{z=\infty } q(-z)\frac{z-1/2} {z-1}q(z-1)\frac{dz}{z} . \nn
\end{align}
It is easy to see that $S_1=0$, $S_4=\frac12 q_0^2$ and $S_2=S_3$. By using 
$$
\frac{z-1/2} {z-1} = 1 + \frac12 \sum_{k\geq 1} z^{-k},\quad z\to \infty, 
$$
we obtain
\begin{align*}
S_3&=\frac 12 \sum_{i\geq0} p_i  \Biggl(\sum_{j=0}^{i+1}  q_j  \binom{i+1}{j} + \frac12 \sum_{j=0}^i  q_j \,\sum_{k=1}^{i+1-j} (-1)^k \binom{i+1}{j+k} \Biggr)\nn\\
&=\frac14 \sum_{i\geq 0} p_i \sum_{j=0}^i  \binom{i}{j} \bigl(2\,  q_{j+1}+  q_j  \bigr). 
\end{align*}
Note that in the last equality we have used the following elementary identity 
$$
\sum_{k=1}^{i+1-j} (-1)^k \binom{i+1}{j+k} = - \binom{i}{j} .
$$
As a result, we proved the formula~\eqref{wl1}. 

To prove the formula~\eqref{wl2}, we first use the identity 
$$
\frac{\Phi(z)}{\Phi(z-2)} = \frac{\left(z-\frac12\right)\left(z-\frac32\right)}{(z-1)\sqrt{z(z-2)}}
$$
to represent~$\Phi_2(z)$ in the form
\[
\Psi_2=z^{3/2} \, (z-2)^{-1/2} \, \frac{\Phi(z)}{\Phi(z-2)} \, e^{-2\p_z}
=\frac{ z\, (z-1/2) \, (z-3/2)}{(z-1) \, (z-2)}   \, e^{-2\p_z}.
\]
Then by calculating the following residue 
$$
-\frac12 \, {\rm Res}_{z=\infty }  \,f(-z) \, \frac{ z \, (z-1/2) \, (z-3/2)}{(z-1) \, (z-2)} \, f(z-2) \,\frac{dz}{z^2}
$$
we arrive at the proof of validity of the formula~\eqref{wl2}. The theorem is proved.
\end{prf}

\section{Virasoro constraints for the GUE partition function with even couplings}\label{virgue}

We show in this section that the GUE partition function $Z_{\tiny\rm even}(x,{\bf s};\e)$ with even couplings satisfies certain Virasoro constraints, and that it also gives a tau function of the discrete KdV hierarchy.  To this end, we first consider the GUE partition function which also depends on the odd integer numbered coupling constants $s_1, s_3, \dots$. We recall the well-known results that this GUE partition function satisfies the Virasoro constraints, and it gives a tau function of the Toda lattice hierarchy. We then take the odd coupling constants to be zero to obtain the desired properties of the GUE partition function with even couplings.

We use the same symbol $Z_N({\bf s})$ as we do in \eqref{part} to denote
the GUE partition function of size $N$ with couplings ${\bf s}=(s_1, s_2, s_3, \dots)$, i.e.
\beq\label{part-GUE}
Z_N({\bf s})=\frac{(2\pi)^{-{N}}}{\rm{Vol}(N)} \int_{{\mathcal H}(N)} e^{- \frac1{\e}\tr V(M;\, {\bf s})} dM
\eeq
with 
$$ V(M;{\bf s})=\frac12 M^2 -\sum_{j\geq 1} s_{j} M^{j}. $$
Introducing the variable~$x$ as we do in~\eqref{zh-36}
and expanding the free energy $\mathcal{F}_N({\bf s}) := \log Z_N({\bf s})$ in powers of $\epsilon$ yields the GUE free energy
\beq\label{genusF-even-odd}
\Fgue(x, {\bf s}; \epsilon):=\left.\mathcal{F}_N({\bf s}) \right|_{N=\frac{x}{\epsilon}} -\frac1{12}\log \epsilon 
= \sum_{g\geq 0} \epsilon^{2g-2} \mathcal{F}_g (x, {\bf s}) .
\eeq
It can be represented in the form \cite{BIZ, thooft1, thooft2, mehta}
\begin{align}
&  \Fgue(x, {\bf s};\epsilon ) = \frac{x^2}{2\epsilon^2} \biggl( \log x -\frac32\biggr) -\frac1{12} \log x +\zeta'(-1) +\sum_{g\geq 2} \epsilon^{2g-2} \frac{B_{2g}}{2g(2g-2)x^{2g-2}} \nn\\
& \qquad + \sum_{g\geq 0} \epsilon^{2g-2} \sum_{k\geq 0} \sum_{i_1, \dots, i_k\geq 1} 
a_g(i_1, \dots, i_k) s_{i_1} \dots s_{i_k} x^{2-2g -\bigl(k-\frac{|i|}2\bigr)}, \label{ribb1} \\
& a_g(i_1, \dots, i_k) = \sum_\Gamma  \frac{1}{\#\, {\rm Sym} \, \Gamma},  \label{ribb2}
\end{align}
where the last summation is taken over all connected oriented ribbon graphs 
of genus $g$ with $k$ vertices of valencies $i_1$, $\dots$, $i_k$.
The exponential 
\beq
e^{\Fgue(x, {\bf s};\epsilon )} \,=:\, \Zgue(x,{\bf s}; \epsilon)
\eeq
is called the GUE partition function.
From \eqref{ribb1} we see that the GUE free energy $\Fgue(x, {\bf s}; \epsilon)$ lives in
 the following Bosonic Fock space
 $$\mathcal{B} = \frac1{\e^2}\mathbb{C}[\epsilon][[x-1,s_1,s_2,\dots]] .$$

By using the shift operator 
\begin{equation}
\Lambda=e^{\e \p_x},
\end{equation}
we define two functions 
\begin{align} 
&U= U(x,{\bf s}; \e) :=  (\lm-1) (1-\lm^{-1}) \Fgue(x, {\bf s};\epsilon ), \label{uvgue} \\
&V =V(x, {\bf s}; \e):=\e \frac{\p}{\p s_1} (\lm-1)\Fgue(x, {\bf s};\epsilon ),\label{uvgue2}
\end{align}
and the operator
\begin{equation}\label{zh-31}
L = \Lambda+ V+ \exp(U) \Lambda^{-1} .
\end{equation}
\begin{lem} \label{guetodalem}
The functions $U, V$ satisfy the following equations of the Toda Lattice hierarchy:
\beq\label{todapol}
\epsilon \frac{\pal L}{\pal s_j} = \left[A_j,  L\right], \quad A_j :=\left( L^j \right)_+ ,\quad \forall\, j\geq 1. 
\eeq
Moreover, $\Zgue$ is a tau function (cf. Definition 1.2.4 in~\cite{DY1}) of the Toda lattice hierarchy.
\end{lem}
The proof of the above lemma can be obtained by using the orthogonal polynomial technique~\cite{mehta}, 
see for example \cite{gmmmo} (see also~\cite{DY1}, esp. Corollary~A.2.2 and Definition~1.2.4 therein).
The following lemma is also well known, see e.g. \cite{gmmmo, MMMM, mehta,  Moro}.

\begin{lem} \label{GUE-toda-virasoro}
The GUE partition function $\Zgue$ satisfies the following Virasoro constraints:
\beq \label{vira-gue0}
L_m^{\rm Toda}({\bf s};\e) \Zgue(x,{\bf s};\e) = 0 , \quad \forall\, m\geq -1.
\eeq
Here the Virasoro operators $L_m^{\rm Toda}=L_m^{\rm Toda}({\bf s};\e)$ are given by
\begin{align}
& L_{-1}^{\rm Toda}:=\sum_{k\geq 2} k s_k \frac{\pal}{\pal s_{k-1}}- \frac{\p}{\p s_{1}} + \frac{x  s_1}{\e^2}, \label{vira-gue3}\\
& L_0^{\rm Toda}:= \sum_{k\geq 1}  k s_k \frac{\pal}{\pal s_{k}} + \frac{x^2}{\e^2} -  \frac{\p}{\p s_{2}},\label{vira-gue2}\\
&L_m^{\rm Toda} := \epsilon^2 \sum_{k=1}^{m-1} \frac{\pal^2}{\pal s_k \pal s_{m-k}} + 2  x \frac{\p }{\p s_m}+ \sum_{k\geq 1}  k s_k \frac{\pal}{\pal s_{k+m}} 
-  \frac{\p}{\p s_{m+2}},\quad m\geq 1.\label{vira-gue1}
\end{align}
They satisfy the commutation relations
$$
\left[L_m^{\rm Toda} , L_n^{\rm Toda}\right] = (m-n) L_{m+n}^{\rm Toda},\quad \forall\, m, n\geq -1. 
$$
\end{lem} 
We note that 
one can use the observation of~\cite{Du2} and the results of~\cite{DZ-vira} to give a simple and straightforward proof of this lemma.

By using the formulae \eqref{ribb1}, \eqref{ribb2} and the fact that the total valency of any ribbon graph is an even number, we have the following lemma on the property of the GUE free energy.
\begin{lem} \label{simple-odd}
The GUE free energy $\Fgue$ satisfies the following property: if $k_1+\dots+k_m$ is an odd number, then
$$
\frac{\p^m \Fgue}{\p s_{k_1} \dots \p s_{k_m}}\bigg|_{s_1=s_3=s_5=\dots=0} \, \equiv \, 0 .
$$
\end{lem}
We also list some other useful identities on the GUE free energy in the next lemma.
\begin{lem}\label{todaidentities}
The following formulae hold true for the GUE free energy $\Fgue$:
\begin{align}
\e^2\frac{\p^2\Fgue}{\p s_1 \p s_1} =& \exp(U), \label{tdd1}\\
\e^2\frac{\p^2\Fgue}{\p s_1 \p s_3} =& \exp(U)\left(V(x)^2+V(x-\e)^2+V(x) V(x-\e)\right) \nn\\
&+ \exp(U) \left(1+ \Lambda+\Lambda^{-1}\right)\exp(U), \label{tdd3}\\
\e^2\frac{\p^2\Fgue}{\p s_2 \p s_2} =& \exp(U) \left(V(x-\e)+ V(x)\right)^2+ \exp(U)\left(\lm+\lm^{-1}\right) \exp(U), \label{tdd4}\\
\e \left(\frac{\p \Fgue}{\p s_2}\right) =& (\Lambda-1)^{-1}\left(V^2  + (\lm+1) \exp(U)\right). \label{td-2}
\end{align}
\end{lem}
\begin{prf}
We can obtain these identities by using (1.2.8) and (2.1.7)--(2.1.13) of \cite{DY1}. The lemma is proved. 
\end{prf}

Now let us consider the GUE free energy and partition function with even couplings, they are obtained from 
the GUE free energy and partition function by putting $s_1=s_3=s_5=\dots=0$, namely,
\begin{align}
& \F_{\tiny\rm even}(x,{\bf s}; \e) = \Fgue(x,s_1=0, s_2, s_3=0,s_4, \dots; \e) ,\nn\\
& Z_{\tiny\rm even}(x,{\bf s}; \e)  = \Zgue(x,s_1=0, s_2, s_3=0,s_4, \dots; \e) \nn
\end{align}
Here and in what follows we restore the notation
$${\bf s}=(s_2,s_4,s_6,\dots)$$
introduced in Sect.\ref{section1}. 
It follows from Lemma \ref{simple-odd} and the definition \eqref{uvgue2} of $V$ that 
after the restriction $s_1=s_3=s_5=\dots=0$ we have
$V \equiv 0$.
The Lax operator $L$ given in \eqref{zh-31} now becomes
\begin{equation}
L= \Lambda+\exp(U) \Lambda^{-1}\label{zh-19} 
\end{equation}
with
\begin{equation}\label{zh-20}
U= U(x,{\bf s};\e)=  (\lm-1) \left(1-\lm^{-1}\right) \F_{\tiny\rm even}(x,{\bf s};\e).
\end{equation}
Here we note that 
$$
\F_{\tiny\rm even}(x,{\bf s};\e)\in \mathcal{B}^{\rm even} = \frac1{\e^2}\mathbb{C}[\epsilon][[x-1,s_2,s_4,\dots]] , \quad U(x,{\bf s};\e) \in \e^2 \mathcal{B}^{\rm even} .
$$

From Lemma~\ref{guetodalem} we have the following lemma.
\begin{lem}\label{zh-18}
The function~$U$ satisfies the discrete KdV hierarchy (aka the Volterra hierarchy)
\begin{equation}\label{dkdv}
\ve \frac{\p L}{\p s_{2k}} = \biggl[\left(L^{2k}\right)_+, \,L\biggr],\quad k\ge 1
\end{equation}
and the initial condition 
\beq \label{inilogx}
\exp(U(x,{\bf 0};\e)) = x .
\eeq
\end{lem}

It should be noted that solution to the equation~\eqref{dkdv} with initial condition~\eqref{inilogx} exists and is unique in~$\e^2 \mathcal{B}^{\rm even}$.
Moreover, one can easily get an analogue of the definition of tau function of 
the discrete KdV hierarchy from~\cite{DY1} such that $Z_{\tiny\rm even}$ is a particular tau function. 
The tau function of any solution to the discrete KdV hierarchy is uniquely determined up to a linear function
of~${\bf s}$ and~$x$. This linear function can further be fixed by the so-called string equation (see below) up to a linear function in~$x$. We omit 
the details because these are just specializations of the results of~\cite{DY1} to the even couplings. 

\begin{emp} 
The $k=1$ flow of the discrete KdV hierarchy \eqref{dkdv} is given by
$$
\frac{\p U}{\p s_2} = \frac1{\e} \left(\Lambda-\Lambda^{-1}\right)\exp(U).
$$
\end{emp}

\begin{theorem} \label{thm-newtau-vira}
Let us introduce a modification $\widetilde{Z}(x,{\bf s};\e)$ of the GUE partition function with even couplings by using the relation
$$
\log Z_{\tiny\rm even}(x,{\bf s};\e) = \left(\lm^{1/2}+\lm^{-1/2}\right) \log \widetilde Z(x,{\bf s};\e).
$$
Then $\widetilde Z(x,{\bf s};\e)$ satisfies the followings system of Virasoro constraints:
\begin{equation}\label{vir-ZZ-dkdv}
L_n^{\rm even} \left( \epsilon^{-1}x, \epsilon^{-1} \tilde{\bf s}, \e \p/\p{\bf s}\right) \widetilde Z(x,{\bf s};\e) = 0,\quad n\geq 0,
\end{equation}
where $\tilde{\bf{s}}=(\tilde{s_2}, \tilde{s_4},\dots)$ is defined by
\begin{equation}\label{zh-2}
\tilde s_{2k} = s_{2k} - \tfrac12 \delta_{k,1},
\end{equation}
and the Virasoro operators $L_n^{\rm even}=L_n^{\rm even} \bigl( \epsilon^{-1}x, \epsilon^{-1}{\bf s}, \e \p/\p{\bf s}\bigr)$ have the expressions 
\begin{align}
&L_0^{\rm even} = \sum_{k\ge 1} k s_{2k}\frac{\p}{\p s_{2k}}+\frac{x^2}{4\ve^2} - \frac1{16}, \label{vire0} \\ 
& L_n^{\rm even} = \ve^2\sum_{k=1}^{n-1}  \frac{\p^2}{\p s_{2k} \p s_{2n-2k}}+ 
x \frac{\p}{\p s_{2n}}+ \sum_{k\ge 1} k s_{2k} \frac{\p}{\p s_{2k+2n}},\quad n\geq 1 . \label{viren}
\end{align}
These operators satisfy the commutation relations
\beq\label{even-comm}
\bigl[L_m^{\rm even}, L_n^{\rm even}\bigr] = (m-n) L^{\rm even}_{m+n}, \quad \forall\,m,n\geq 0 .
\eeq
\end{theorem}
\begin{prf}
The Virasoro commutation relation \eqref{even-comm} can be verified straightforwardly.
It then suffices to prove \eqref{vir-ZZ-dkdv} for $n=0,1,2$, because the rest of \eqref{vir-ZZ-dkdv} can be proved by using
 \eqref{even-comm}. We denote $\widetilde \F= \log \widetilde Z(x,{\bf s};\e)$.

Let us start with the case when $n=0$. By taking $m=0$ in \eqref{vira-gue0} of Lemma \ref{GUE-toda-virasoro} we obtain 
$$
\sum_{k\geq 1}  k s_k {\pal \Zgue \over \pal s_{k}} + \frac{x^2}{\e^2} \Zgue -  \frac{\p \Zgue}{\p s_{2}} = 0 .
$$
Put $s_1=s_3=s_5=\dots=0$ in this equation and divide it by $Z_{\tiny\rm even}$ we arrive at 
\beq \label{v02}
\sum_{k\geq 1}  2k s_{2k} {\pal \F_{\tiny\rm even} \over \pal s_{2k}} 
+ \frac{x^2}{\e^2}  -  \frac{\p \F_{\tiny\rm even}}{\p s_{2}} = 0 .
\eeq
Now by applying the operator $\left(\lm^{1/2}+\lm^{-1/2}\right)^{-1}$ on both sides of~\eqref{v02} we get
$$
\sum_{k\geq 1}  k\, s_{2k} \, {\pal \widetilde \F \over \pal s_{2k}} + \frac{x^2}{4\e^2} - \frac1{16} - \frac12\, \frac{\p \widetilde{\F} }{\p s_{2}} = 0 .
$$
This proves~\eqref{vir-ZZ-dkdv} with $n=0$. 

For the case when $n=1$, we take $m=2$ in~\eqref{vira-gue0} and obtain
$$
2\,x\,\frac{\p \Zgue}{\p s_2}+\sum_{k\geq 1}  k\, s_k \, {\pal \Zgue \over \pal s_{k+2}} 
+ \epsilon^2 \, {\pal^2 \Zgue \over \pal s_1^2} -  \frac{\p \Zgue}{\p s_{4}} = 0 .
$$
Put $s_1=s_3=s_5=\dots=0$ in this identity and divide it by $Z_{\tiny\rm even}$ we 
get 
$$
2\,x\,\frac{\p \F_{\tiny\rm even}}{\p s_2}
+\sum_{k\geq 1}  2k\, s_{2k} \, {\pal \F_{\tiny\rm even} \over \pal s_{2k+2}} +  
\epsilon^2 \, \Biggl({\pal^2 \Fgue \over \pal s_1^2}+ \biggl({\pal \Fgue \over \pal s_1}\biggr)^{\!\!2} \, \Biggr)\Bigg|_{s_1=s_3=\dots=0} 
-  \frac{\p \F_{\tiny\rm even}}{\p s_{4}} = 0 .
$$
Then from Lemma~\ref{simple-odd} and the identity~\eqref{tdd1} it follows that
\beq\label{vir-ZZ-dkdv2}
2 x \frac{\p \F_{\tiny\rm even}}{\p s_2}
+\sum_{k\geq 1}  2 k  s_{2k} {\pal \F_{\tiny\rm even} \over \pal s_{2k+2}} + \exp(U) -  \frac{\p \F_{\tiny\rm even}}{\p s_{4}} = 0.
\eeq
On the other hand, by putting $s_1=s_3=s_5=\dots=0$ in  \eqref{td-2} we obtain
\beq \label{simple1}
 \e \frac{\lm-1}{\lm+1} \, \left(\frac{\p \F_{\tiny\rm even}}{\p s_2}\right) =\exp(U).
\eeq
By applying the operator
$\left(\lm^{1/2}+\lm^{-1/2}\right)^{-1}$ on both sides of \eqref{vir-ZZ-dkdv2} and by using \eqref{simple1} we arrive at
$$
x \, \frac{\p \widetilde\F}{\p s_{2}}+ \sum_{k\ge 1} k\, s_{2k} \, \frac{\p \widetilde\F}{\p s_{2k+2}}  - \frac12 \frac{\p \widetilde\F}{\p s_{4}} = 0,
$$
which proves the Virasoro constraint \eqref{vir-ZZ-dkdv} for $n=1$. 

The validity of the Virasoro constraint \eqref{vir-ZZ-dkdv} for $n=2$ can be proved
in a similar way as we did for the $n=1$ case by 
using Lemmas \ref{GUE-toda-virasoro}--\ref{todaidentities}, so we omit the details here. The theorem is proved.
\end{prf}

\begin{rmk}
It was an open question to find the Virasoro constraints for $Z_{\tiny\rm even}(x,{\bf s})$ in a compact form \cite{MMMM,gmmmo}.
Of course, $Z_{\tiny\rm even}$ \textit{itself} does satisfy certain Virasoro type constraints, 
but these constraints may contain non-linear terms. For example, the $L_1^{\rm even}$ constraint for $Z_{\tiny\rm even}$ reads
$$
\Biggl(2 x \frac{\p }{\p s_2}
+\sum_{k\geq 1}  2k s_{2k}  \frac{\pal}{\pal s_{2k+2}} +\exp(U)-  \frac{\p }{\p s_{4}}\Biggr) Z_{\tiny\rm even}= 0, 
\quad U=  \left(\lm-1\right) \left(1-\lm^{-1}\right)  \log Z_{\tiny\rm even}, 
$$
which is a non-linear action on $Z_{\tiny\rm even}$. 
The {\em key} of our study is the introduction of the modification $\widetilde Z$ 
of the GUE partition function with even couplings which linearizes the nonlinear constraints, 
and this enables us to write down all the Virasoro constraints in a closed form. 
\end{rmk}

To finish this section, let us note that the Virasoro constraints~\eqref{vir-ZZ-dkdv} 
correspond to the Virasoro symmetries of the discrete KdV hierarchy. The first two of them,
in terms of  the function $U$ defined in~\eqref{zh-20}, have the following form:
\begin{align}
& \frac{\p U}{\p \tau_0} = \sum_{k\geq 1} k s_{2k} \frac{\p U}{\p s_{2k}}+1,  \nn\\
& \frac{\p U}{\p \tau_1} = \frac12 \left(3\lm+3\lm^{-1}+2\right) \exp(U)+ x \frac{\p U}{\p s_2}+\sum_{k\ge 1} k s_{2k} \frac{\p U}{\p s_{2k+2}}. \nn
\end{align}

\section{Proof of the Main Theorem}\label{section4}
In the previous section we introduced a modification $\widetilde{Z}(x,{\bf s};\e)$ of the GUE partition function with even couplings, which plays a crucial role in our presentation of the associated Virasoro constraints in terms of linear actions of certain Virasoro operators on the partition function.  We also introduced the following modification of the GUE free energy with even couplings
\begin{equation}
\widetilde \F(x,{\bf s};\e) = \Bigl(\Lambda^{1/2}+\Lambda^{-1/2}\Bigr)^{-1} \F_{\tiny\rm even}(x,{\bf s};\e).\label{definingwf}
\end{equation}
From Theorem \ref{thm-newtau-vira} we know that $\widetilde \F(x,{\bf s};\e)$
satisfies the Virasoro constraints
\begin{equation} \label{levenwfagain}
L_n^{\rm even}\left(\e^{-1} x, \e^{-1} \tilde{\textbf{s}},  \e \partial /\partial \textbf{s}\right) e^{ \widetilde \F(x,{\bf s};\e)}= 0,\quad n\geq 0,
\end{equation}
where the linear operators $L_n^{\rm even}$ are given by \eqref{vire0}, \eqref{viren}.
Now let us introduce the following modification of the special cubic Hodge free energy:
\begin{equation}
\widehat \F(x,{\bf s};\e)=
\H_{\tiny\rm cubic}\left(\bt(x,{\bf s}); \sqrt{2}\e\right) + \e^{-2} \, \frac{A(x, {\bf s})}2 + \frac{\zeta'(-1)}2 . \label{def2wfs}
\end{equation}
Here $A(x, {\bf s})$ and ${\bf t}(x, {\bf s})$ are defined in \eqref{axs} and \eqref{time-sub}. 

In order to prove the Main Theorem, we first show that $\widehat \F(x,{\bf s};\e)$
satisfies the same system of Virasoro constraints as $\widetilde \F(x,{\bf s};\e) $ does, and then by using the genus expansions
of the special cubic Hodge free energy and the GUE free energy with even couplings we arrive at the identity
\begin{equation}\label{zh-3}
\widetilde \F(x,{\bf s};\e)=\widehat \F(x,{\bf s};\e).
\end{equation} 
In this way we prove the validity of~\eqref{id1} and also the Main Theorem.

The following lemma is important in establishing the relationship between the Virasoro constraints \eqref{definingwf} and the ones for $\widehat \F(x,{\bf s};\e)$.
\begin{lem} \label{keyintro} 
The Virasoro operators $\widetilde L_m^{\rm cubic}$, $L_m^{\rm even}$ defined in 
\eqref{wl0}--\eqref{wl2} and in \eqref{vire0}, \eqref{viren} for $m\ge 0$ are related by the following identity:
\begin{equation}
e^{-\frac{A(x, {\bf s})}{2\e^2} } \circ L_m^{\rm even} \left( \epsilon^{-1}x, \epsilon^{-1} \tilde{\bf  s}, \e \,\p/\p{\bf  s}\right)\circ e^{\frac{A(x, {\bf s})}{2\e^2} } 
= 4^m   \widetilde L_m^{\rm cubic}\left((\sqrt{2} \e)^{-1}\tilde{\bf t}, \sqrt{2}\e \,\p/\p{\bf t}\right) \Big|_{\small {\bf t} = {\bf t}(x,{\bf s})}.
\label{deflmeven}
\end{equation}
Here ${\bf t}(x, {\bf s})=(t_0(x, {\bf s}),  t_1(x, {\bf s}), \dots)$ is defined as in~\eqref{time-sub}.
\end{lem}
\begin{prf}
In the case when $m=0$, we have
\begin{align*}
&e^{-\frac{A(x, {\bf s})}{2 \e^2}} \circ L_0^{\rm even} \left(\e^{-1} x, \e^{-1} \tilde{\bf{s}}, \e \, \partial /\partial {\bf{s}}\right) \circ e^{\frac{A(x, {\bf s})}{2 \e^2}}\\
&=L_0^{\rm even} \left(\e^{-1} x, \e^{-1} \tilde{\bf{s}}, \e\, \partial /\partial \textbf{s}\right)-\frac{1}{2\e^2} \left[A(x, {\bf s}), L_0^{\rm even} \left(\e^{-1} x, \e^{-1} \tilde{\bf{s}}, \e\, \partial /\partial \textbf{s}\right)\right]\\
&=\sum_{k\ge 1} k {\bar{s}}_{k}\frac{\p }{\p \bar{s}_{k}}-\frac{\p}{\p \bar{s}_1}+\frac{x^2}{4\ve^2} - \frac1{16}
+ \frac{1}{4 \e^2} \sum_{k_1, k_2\ge 1} k_1 k_2 {\bar{s}}_{k_1} {\bar{s}}_{k_2}\\
&\quad +
\frac{x-1}{2 \e^2} \sum_{k\ge 1} k {\bar{s}}_{k}+\frac{1}{2 \e^2}\left(\frac12-x\right)\\
&=\sum_{i\ge 1} t_i \frac{\p}{\p t_{i-1}}-\frac{\p}{\p t_0}+\frac{x^2}{4\ve^2} - \frac1{16}
+\frac{1}{4 \e^2}(t_0+1-x)^2\\
&\quad +\frac{x-1}{2 \e^2} (t_0+1-x)+\frac{1}{2 \e^2}\left(\frac12-x\right)\\
&=\sum_{i\ge 1} t_i \frac{\p}{\p t_{i-1}}-\frac{\p}{\p t_0}+\frac{t_0^2}{4\ve^2} - \frac1{16}\\
&= \widetilde L_0^{\rm cubic} \left(\bigl(\sqrt{2}\e\bigr)^{-1}\tilde{\bf t}, \sqrt{2}\e\p/\p{\bf t}\right).
\end{align*}
In a similar way, one can prove the validity of~\eqref{deflmeven} for $m=1, 2$. 
By using the Virasoro commutation relations~\eqref{zh-6} and~\eqref{even-comm} we know that 
the identity~\eqref{deflmeven} also holds true for $m\ge 3$. The lemma is proved.
\end{prf}

From Theorem~\ref{vira-2-lem} and the above lemma we obtain the following proposition.
\begin{prp} \label{virasorowfstarthm}
The function $\widehat{\F}(x,{\bf s};\e)$ satisfies the following Virasoro constraints
\beq  \label{virasoroevenfstar}
L_n^{\rm even} \left( \epsilon^{-1}x, \epsilon^{-1} \tilde {\bf s}, \e\, \p/\p{\bf s}\right) e^{ \widehat {\F}(x,{\bf s};\e)}= 0,\quad n\geq 0. 
\eeq
\end{prp}

To deduce the validity of the identity \eqref{zh-3} from the Virasoro constraints \eqref{levenwfagain} and \eqref{virasoroevenfstar}, we need to use the property of the following genus expansions of  the special cubic Hodge free energy $\H_{\tiny\rm cubic}(\bt;\e)$ and the GUE free energy $\F_{\tiny\rm even}(x,{\bf s};\e)$ 
with even couplings:
\begin{align}
&  \H_{\tiny\rm cubic}(\bt;\e) = \sum_{g=0}^\infty \e^{2g-2}\, \H_g(\bt) ,  \label{genushhh}\\
&  \F_{\tiny\rm even}(x,{\bf s};\e) = \sum_{g=0}^\infty \e^{2g-2}\, \F_g(x,{\bf s}). \label{genusfff}
\end{align}
Recall that the genus $0, 1$ parts of Conjecture \ref{conjecture1} is proved in \cite{DY2}, i.e. we have
\begin{align}
& \F_0(x, {\bf s}) = \H_0(\bt(x,{\bf s})) + A(x, {\bf s}), \label{g0p}\\
& \F_1(x, {\bf s}) = 2 \, \H_1(\bt(x,{\bf s})) + \frac18 \, \frac{\p^2 \H_0(\bt(x,{\bf s}))}{\p x^2}+ \zeta'(-1).\label{g1p}
\end{align}
From the definitions given in \eqref{definingwf} and \eqref{def2wfs} it follows that $\widetilde \F$ and $\widehat \F$ also have the genus expansions
\begin{align}
&\widetilde \F(x,{\bf s};\e) =:  \sum_{g=0}^\infty \e^{2g-2}\, \widetilde \F_g(x,{\bf s}) ,  \label{genuswf}\\
&  \widehat \F(x,{\bf s};\e) =:\sum_{g=0}^\infty \e^{2g-2}\, \widehat \F_g(x,{\bf s}),\label{genuswfstar}
\end{align}
and we have
\begin{align}
&\widetilde \F_0(x,{\bf s}) = \frac{\F_0(x,{\bf s})}2 ,  
\quad \widehat \F_0(x,{\bf s}) = \frac{\H_0\left(\bt(x,{\bf s})\right)}2 + \frac {A(x, {\bf s})}{2},\\
&\widetilde \F_1(x,{\bf s}) = \frac12 \F_1(x,{\bf s})-\frac1{16} \frac{\p^2 \F_0(x,{\bf s})}{\p x^2},\quad \widehat \F_1(x,{\bf s}) = \H_1\left(\bt(x,{\bf s})\right) +\frac{\zeta'(-1)}2.
\end{align}
Thus by using~\eqref{g0p} and~\eqref{g1p} we 
obtain the following identities 
\beq\label{a1111}
\widetilde \F_0(x,{\bf s}) = \widehat \F_0(x,{\bf s}),\quad \widetilde \F_1(x,{\bf s}) = \widehat \F_1(x,{\bf s}).
\eeq

Let us recall two lemmas on properties of the genus $g$ free energies $\mathcal{H}_g(\bt)$ and $\F_g(x, {\bf s})$ which are proved in \cite{DLYZ} and \cite{DY2}. 

\begin{lem}[\cite{DLYZ}] \label{polyHodge} 
There exist functions $H_g(z,z_1, z_2, \dots, z_{3g-2})$, $g\geq 1$ of independent variables $z$, $z_1$, $z_2$, $\dots$ such that
\beq\label{quasiH}
\mathcal{H}_g(\bt)= H_g\biggl( v(\bt ), \frac{\pal v(\bt )}{\pal t_0}, \dots, \frac{\pal^{3g-2} v(\bt)}{\pal t_0^{3g-2}}\biggr), \quad g\geq 1
\eeq
and that 
$$
H_1=\frac{1}{24}\log z_1-\frac{1}{16}z,\quad \sum_{j=1}^{3g-2} j \, z_j \frac{\p H_g}{\p z_j} = (2g-2) \, H_g, \quad g\ge 2.
$$
Here $v({\bf t}):=\frac{\p^2 \H_0({\bf t})}{\p t_0^2}$ is the unique power series solution to  
\beq\label{v-dispkdv-intro-b}
v = t_0+ \sum_{i\geq 1} t_i \frac{v^i}{i!},\quad v(\bt)|_{t_i=0,\, i\ge 1}=t_0.
\eeq
\end{lem}
We note that the explicit expression of~$v(\bt)$ in this lemma is given in~\eqref{v-dispkdv-intro}.

\begin{lem} [\cite{DY2}] \label{Fgthm}
There exist functions $F_g(z,z_1, \dots, z_{3g-2})$, $g\geq 1$ of independent variables 
$z$, $z_1$, $z_2$, $\dots$ such that
\beq\label{quasiF}
\F_g(x, {\bf s}) = F_g \biggl( u(x, {\bf s}), \frac{\pal u(x, {\bf s})}{\pal x}, \dots, \frac{\pal^{3g-2}u(x, {\bf s})}{\pal x^{3g-2}}\biggr), \quad g\geq 1,
\eeq
and that 
$$
F_1=\frac{1}{12}\log z_1+\frac{i \pi}{24}+\zeta'(-1),\quad\sum_{j=1}^{3g-2} j \, z_j \frac{\p F_g}{\p z_j} = (2g-2) \, F_g,\quad g\ge 2.
$$
Here
$u(x,{\bf s}):=\frac{\pal^2 \F_0(x,{\bf s})}{\pal x^2}=\log w(x,{\bf s})$, and 
$w(x,{\bf s})$ is the unique series solution to
\beq\label{weven}
w = x+ \sum_{k\geq 1} \, k\, \bar s_k  \, w^k,\qquad w(x,{\bf s})|_{\bar{s}_k=0,\, k\ge 1}= x.
\eeq
\end{lem}

We note that the explicit expression for~$w(x,{\bf s})$ in this lemma is given by 
$$
w(x,{\bf s})=\sum_{n=1}^\infty \frac1{n} \sum_{i_1,\dots,i_n\geq 0 \atop i_1+\cdots + i_{n}=n-1} {\rm wt}(i_1)\cdots {\rm wt}(i_n) \, \bar s_{i_1}\cdots \bar s_{i_n},
$$
where we put $\bar s_0=x$ and denote
$$
{\rm wt}(i)= \begin{cases} 
1, & i=0\\
i,  & {\rm otherwise.}\end{cases} 
$$

\begin{lem} \label{wfwfstar}
For any $g\geq 1$, there exist functions $\widetilde F_g(z,z_1,\dots,z_{3g-2})$ and $\widehat F_g(z,z_1,\dots,z_{3g-2})$ of independent 
variables $z,z_1,\dots,z_{3g-2}$ such that
\begin{align}
& \widetilde \F_g(x,{\bf s}) = \widetilde F_g\biggl( u(x,{\bf s}), \frac{\pal u(x, {\bf s})}{\pal x}, \dots,  \frac{\pal^{3g-2}u(x, {\bf s})}{\pal x^{3g-2}}\biggr),  \label{quasiwf}\\
& \widehat \F_g(x,{\bf s}) = \widehat F_g \biggl( u(x,{\bf s}), \frac{\pal u(x, {\bf s})}{\pal x}, \dots, \frac{\pal^{3g-2}u(x, {\bf s})}{\pal x^{3g-2}}\biggr) \label{quasiwfstar}
\end{align}
with $u(x,{\bf s})$ defined as in Lemma \ref{Fgthm}.
\end{lem}
\begin{prf}
Observe that, as in~\cite{DY2}, under the substitution 
$$t_i(x, {\bf s}) = \sum_{k\geq 1} k^{i+1} \bar{s}_k  - 1 + \delta_{i,1} + x \, \delta_{i,0} , $$
we have $v(\bt(x,{\bf s}))= u(x,{\bf s})$, where $v(\bt)$ is defined as in Lemma \ref{polyHodge}. 
The lemma then follows from Lemmas \ref{polyHodge}, \ref{Fgthm} and the defining equations \eqref{definingwf}, \eqref{def2wfs}.
\end{prf}

Now let us proceed to show that equation~\eqref{virasoroevenfstar} (or~\eqref{levenwfagain}) possesses a unique solution~$\widetilde{\F}$ under a certain assumption.

\begin{prp}\label{zh-11}
Consider the Virasoro constraints
\beq  \label{zh-7}
L_n^{\rm even} \left( \epsilon^{-1}x, \epsilon^{-1} \tilde {\bf s}, \e\, \p/\p{\bf s}\right) e^{ \PP(x,{\bf s};\e)}= 0,\quad n\geq 0
\eeq
for function $\PP$ of the form
\begin{equation}\label{zh-8}
\mathcal{\PP}=\e^{-2} \mathcal{P}_0(x, {\bf s})+\triangle P,\quad
\triangle P=\sum_{g\ge 1} \e^{2g-2} P_g\biggl( u(x,{\bf s}), \frac{\pal u(x, {\bf s})}{\pal x}, \dots,  \frac{\pal^{3g-2}u(x, {\bf s})}{\pal x^{3g-2}}\biggr),
\end{equation}
where $\PP_0=\widetilde{\F}_0(x, {\bf s})=\frac12 \F_0(x, {\bf s})$, $u(x, {\bf s})=\frac{\pal^2 \F_0(x,{\bf s})}{\pal x^2}$. Then $\triangle P$ satisfies the following equation:
\begin{align}
& \sum_{k\geq 0} 
\biggl(\p_x^k \left(\frac1{B^2}\right) + \sum_{r=1}^k \binom{k}{r} \, \p_x^{r-1} \left(\frac1B\right) \; \p_x^{k-r+1} \left(\frac1B\right)\biggr)  
\frac{\p \triangle P}{\p u_k} \nn\\
& -\e^2 \sum_{k_1, k_2} \p_x^{k_1+1} 
\left(\frac1B\right) \p_x^{k_2+1} \left(\frac1B\right)  \biggl(\frac{\p \triangle P}{\p u_{k_1}}  \frac{\p \triangle P}{\p u_{k_2}}  + \frac{\p^2 \triangle P}{\p u_{k_1} \p u_{k_2}}\biggr)  \nn\\
& - \e^2 \, \sum_{k} \p_x^{k+2} \left(\frac1{8B^4} - \frac1{4B^2} \right) \, 
\frac{\p \triangle P}{\p u_k} 
+ \frac1 {8B^2} - \frac1{16B^4} = 0 .   \label{loopf}
\end{align}
Here we denote $B=\sqrt{1-\frac{4e^u}{\lambda}}$ and $u_k=\frac{\p^k u(x, {\bf s})}{\p x^k}$ with $u_0=u$. Note that equation~\eqref{loopf} holds true identically in~$\lambda$.
\end{prp}
\begin{prf}
Let us denote 
\[S(\lambda):=\sum_{k\geq 1} k (s_{2k}-\frac12 \delta_{k,1}) \lambda^{k-1}, \quad T(\lambda) := \sum_{\ell\geq 1}\frac1{\lambda^{\ell+1}} \frac{\p}{\p {s_{2\ell}}}.\]
Then it is easy to check that the Virasoro constraints~\eqref{zh-7} can be represented as 
\begin{align}
&\left(S(\lambda) T(\lambda) \right)_{\leq -2} \PP_0  
+ \frac x\lambda  T(\lambda) \PP_0 
+ \left(T(\lambda) \PP_0 \right)^2 + \frac{x^2}{4 \, \lambda^2} = 0,\label{genus0const} \\
&\left(\left(S(\lambda) T(\lambda) \right)_{\leq -2}  + \frac x\lambda \, T(\lambda)
+  \e^2 \, T(\lambda)^2
+ 2 \, \left(T(\lambda) \PP_0\right) T(\lambda)\right)\triangle P \nn\\
 &=\frac{1}{16 \, \lambda^2}-T(\lambda)^2 \PP_0-\e^2 \, \left(T(\lambda) \, \triangle P \right)^2 .  \label{loopeqpre}
\end{align}
Here we use the notion~$(W(\lambda))_{\leq -2}$ to denote the part of a power series~$W(\lambda)$ 
which consists of terms with powers of $\lambda$ less than or equal to~$-2$. 
We note that the equation~\eqref{genus0const} holds true since it is just the genus zero Virasoro constraint for~$\F_0(x, {\bf s})$.

We know from \cite{DY2} that the genus zero GUE two point functions can be represented in terms
of~$u(x, {\bf s})$ as follows:
$$
\frac{\p^2\PP_0}{\p s_{2 k} \p x} = \frac12 \binom{2 k}{k} \, e^{k u},\quad  \frac{\p^2 \PP_0}{\p s_{2 l} \p s_{2k}}= \frac12 \frac{k l}{k+l} \binom{2k}{k} \binom{2 l}{l} e^{(k+l)u},\quad k, l\ge 1.
$$ 
From these identities it follows that
\begin{align}
&T(\lambda) \PP_{0, x}= -\frac1{2\lambda} + \frac1{2\lambda B},\quad
T(\lambda)^2 \, \PP_0 = - \frac{\lambda-8e^u}{16\lambda(\lambda-4e^u)^2} + \frac1{16\lambda^2},\label{zh-9}\\
&T(\lambda) u = \frac1{\lambda}\p_x \biggl(\frac1{ B}\biggr) , 
 \quad  T(\lambda) u_k = \frac1{\lambda}\p_x^{k+1}\biggl(\frac1{ B}\biggr),\quad k\ge 1.   \label{zh-10}
\end{align}
On the other hand, by differentiating the l.h.s of \eqref{genus0const} with respect to~$x$ twice we obtain
\begin{align}
& \frac12 \left(S(\lambda) T(\lambda) \right)_{\leq -2} u
+ \frac 2 \lambda T(\lambda) \PP_{0, x}  + \frac x {2\lambda} T(\lambda) u +2\left(T(\lambda) \PP_{0, x} \right)^2 \nn\\
& \quad 
 + \left(T(\lambda) \PP_0 \right)\left(T(\lambda) u \right)  + \frac{1}{2 \, \lambda^2} = 0 . \nn
\end{align}
Thus by using the relations \eqref{zh-9}, \eqref{zh-10} we obtain
\begin{align}
&2\left(T(\lambda) \PP_0 \right)\left(T(\lambda) u_k \right)
=-\left(S(\lambda) T(\lambda) \right)_{\leq -2} u_k
-\frac x \lambda T(\lambda) u_k-\frac1{\lambda^2} \p_x^{k}\left(\frac1{B^2}\right) \nn\\
&\quad -  \frac1{\lambda^2}\sum_{r=1}^k \binom{k}{r}\p_x^{r-1} \left(\frac1B\right)\p_x^{k-r+1} \left(\frac1B\right),\quad k\ge 0. \label{nonlocal}
\end{align}
Note that when acting on functions of $u_k=\p_x^k u(x, {\bf s}), k\ge 0$, the operators 
$T(\lambda)$ and $S(\lambda)$ have the following properties:
\begin{align}
& T(\lambda) = \frac1{\lambda} \sum_{q\geq 0} 
 \p_x^{q+1} \left( \frac1B\right) \frac{\p }{\p u_q},  \nn \\
&\left( S(\lambda) T(\lambda) \right)_{\leq -2} = 
\sum_{q\geq 0}  \left( S(\lambda)T(\lambda) \right)_{\leq -2} (u_q)  \frac{\p}{\p u_q}, \nn\\
&T(\lambda)^2 = \frac1{\lambda^2} \sum_{q\geq 0} \p_x^{q+2}  \left( \frac1{8 B^4}-\frac1{4B^2}\right) \frac{\p}{\p u_q} 
 + \frac1{\lambda^2} \sum_{p, q\geq 0} \sum_{m, l\geq 1}  \p_x^{q+1} \left(\frac1B\right) \p_x^{p+1}\left(\frac1B\right) \frac{\p}{\p u_p} \frac{\p}{\p u_q}.  \nn
\end{align}
By using these relations and the formulae given in \eqref{zh-9}--\eqref{nonlocal}, we obtain from \eqref{loopeqpre} the equation \eqref{loopf}. 
The proposition is proved.
\end{prf}

Following the notations of~\cite{DZ-norm}, we call equation~\eqref{loopf} the {\em loop equation} for the modified GUE potential.

\begin{prp}\label{zh-12}
The loop equation given by \eqref{loopf} has a unique solution 
\[\{P_g(u, u_1,\dots, g_{3g-2})\,|\, g\ge 1\}\] 
up to the addition of constants.
\end{prp}
\begin{prf}
The right hand side of the loop equation \eqref{loopf} can be expanded as a series
$\sum_{g\ge 1}A_{g} \e^{2g-2}$ in $\e^2$. From the coefficient of $\e^0$ we get the 
equation
\[\frac32 u_1 \frac{\p P_1}{\p u_1}\frac1{B^4}-\frac32 u_1 \frac{\p P_1}{\p u_1}\frac1{B^2}+\frac{\p P_1}{\p u}\frac1{B^2}+\frac1{8 B^2}-\frac1{16 B^4}=0,\]
which is equivalent to the equations
\[
\frac32 u_1\frac{\p P_1}{\p u_1}=\frac1{16},\quad 
- \frac32 u _1 \frac{\p P_1}{\p u_1}+ \frac{\p P_1}{\p u}=-\frac{1}{8}.\]
By solving this system of linear equations we obtain  
\[P_1=\frac1{24}\log{u_1}-\frac1{16} u+\textrm{constant}.\]
In general, from the coefficients of~$\e^{2g-2}$ we get a system of linear equations 
for $\frac{\p P_g}{\p u}, \frac{\p P_g}{\p u_1}, \dots, \frac{\p P_g}{\p u_{3g-2}}$ with upper triangular coefficient matrix~$M_g$. It is easy to see that 
\[
\det M_g= u_1^{(3g-2)(3g-1)/2}\prod_{j=1}^{3g-2} \frac{(2j+1)!!}{2^j}. 
\]
So the loop equation uniquely fixes the gradients of the function~$P_g$, and we prove the uniqueness part of the proposition. 
The existence part of the proposition follows from the result of Lemma~\ref{wfwfstar}. The proposition is proved.
\end{prf}

Now we are ready to finish the proof of the Main Theorem. From~\cite{DY1} we already 
know the identities 
\[\widetilde{\F}_g(x, {\bf s})=\widehat{\F}_g(x, {\bf s}),\quad g=0,1;\]
see~\eqref{a1111}.
We also know from~\cite{DY1} that 
\[
\sum_{j=1}^{3g-2} j u_j \frac{\p \widetilde F_g}{\p u_j} = (2g-2) ,\quad 
\sum_{j=1}^{3g-2} j u_j  \, \frac{\p \widehat F_g}{\p u_j} = (2g-2) \widehat F_g,\quad g\ge 2.
\]
Thus the identity \eqref{zh-3} follows from the Virasoro constraints \eqref{levenwfagain}, 
\eqref{virasoroevenfstar}, the genus expansion property \eqref{genuswf}, \eqref{genuswfstar},
\eqref{quasiwf}, \eqref{quasiwfstar}, and Propositions \ref{zh-11}, \ref{zh-12}.
The Main Theorem is proved. 

\section{Proof of Corollary~\ref{zh-15}} \label{zh-25}
We know from the Main Theorem that, under the substitution \eqref{time-sub}, the following identity holds true:
\beq\label{id-again}
 \mathcal{F}_{\tiny\rm even}(x, {\bf s};\e) = \left(\Lambda^{\frac12}+\Lambda^{-\frac12}\right) \, \H_{\tiny\rm cubic} \left(\bt(x,{\bf s}); \sqrt{2}\e\right) + \e^{-2} \, A(x, {\bf s}) +\zeta'(-1).
\eeq
Here $\F_{\tiny\rm even}$ is the GUE free energy with even couplings, $A(x, {\bf s})$ is defined in \eqref{axs}. 
Then the function $U(x,{\bf s};\e)$ defined by \eqref{zh-20}, i.e.
$$U(x,{\bf s};\e)= (\lm-2-\lm^{-1}) \, \F_{\tiny\rm even}(x,{\bf s};\e),$$ 
satisfies the discrete KdV hierarchy \eqref{dkdv}.
On the other hand, from the definition of $W({\bf t}, \e)$ given in \eqref{zh-16} it follows that
\[U(x, {\bf s}; \e)=W({\bf t}(x, {\bf s});\e).\]
We also know from the relation \eqref{time-sub} between ${\bf t}$ and ${\bf s}$ that
\[\frac{\p}{\p t_0}=\frac{\p}{\p x},\quad \frac{\p}{\p {\bar s}_k}=\sum_{i\ge 0} k^{i+1} \frac{\p}{\p t_i},\quad k\ge 1.\]
Thus $W({\bf t};\e)$ satisfies the discrete KdV hierarchy \eqref{zh-21}.
The corollary is proved.

\section{Conclusion}\label{zh-26}
We prove in this paper the Hodge--GUE correspondence conjecture that is proposed in~\cite{DY2},
and show that the partition function of the special cubic Hodge integrals of the form~\eqref{zh-22} gives a tau function of the discrete KdV hierarchy as it is conjectured in~\cite{DLYZ}.

Let us briefly review an application of the Main Theorem that is presented in~\cite{DY2}. 
Assuming the validity of the Hodge--GUE conjecture the following formula was obtained in~\cite{DY2}:
\begin{align}
& F_g(z, z_1,\dots, z_{3g-2})\nn\\
& = \frac{z_{2g-2}}{2^{2g}(2g)!} + \frac{D_0^{2g-2} [H_1(z; z_1)]}{2^{2g-3} (2g-2)!}
+\sum_{m=2}^g \frac{2^{3m-2g}}{(2g-2m)!} D_0^{2(g-m)} 
\bigl[H_m(z, z_1,\dots, z_{3m-2})\bigr] , \label{now-identity!}
\end{align}
here $D_0:=z_1 \, \p_z+ \sum_{k\geq 1} z_{k+1} \, \p_{z_k}$, $F_g$ and $H_m$ are related to the  GUE and Hodge free energies by \eqref{quasiH}, \eqref{quasiF}, and $g\ge 2$.  Based on this formula and on the algorithm of computing 
$H_g$ developed in~\cite{DLYZ}, 
the explicit formulae for $F_g,\,g=1, \dots, 5$ are obtained in~\cite{DY2} (see Conjectures 3.3.1--3.3.2 therein). 
For any $g\geq 2$, $F_g$ was also explicitly expressed
by primitive special cubic Hodge integrals~\cite{DY2} (see Conjecture~1.5.1 therein). These conjectures are now theorems as they are consequences 
of the Hodge--GUE conjecture proved in the present paper. 
The above mentioned results, obtained from the Hodge--GUE correspondence, produce an algorithm which
allows one to compute the number of tilings of labelled mixed $2m$-gons on a given oriented real surface. The algorithm is in particular quite efficient when the number of polygons is {\it large}\footnote{It is worth mentioning that
in~\cite{DY1} another algorithm for the ribbon graph enumeration, based on the matrix-resolvent method, is obtained which is in particular quite efficient for {\it large genus}.}.

It is interesting to consider the relationship of the special cubic Hodge integrals~\eqref{chbcy} satisfying the local Calabi--Yau condition~\eqref{zh-23},  random matrix models and integrable systems.  In~\cite{DLYZ, LZZ} it is conjectured that the partition function of these Hodge integrals is a tau function of the so-called fractional Volterra hierarchy. We will study the validity of this conjecture in a subsequent publication.

\begin{appendices}
\renewcommand{\thesection}{A}
\renewcommand{\theequation}{A.\arabic{equation}} 
\section{Givental quantization}\label{appa} \par

Denote by $\mathcal{V}$ 
the space of Laurent polynomials in $z$ with coefficients in $\mathbb{C}$.  
Define a symplectic bilinear form $\omega$ on $\V$ by
$$
\omega(f,g) \,:=\, - {\rm Res}_{z=\infty } \, f(-z)\, g(z) \, \frac{dz}{z^2} = - \omega(g,f) , \quad \forall \, f,g\in \H . 
$$

The pair $(\V,\omega)$ is called a Givental symplectic space. For any $f\in \mathcal{V}$, write
$$
f = \sum_{i\geq 0}  \, q_i \, z^{-i} + \sum_{i\geq 0} \, p_i \, (-z)^{i+1} .
$$
Then $\{q_i,\,p_i\}|_{i=0}^\infty$ gives a system of canonical coordinates for $(\V,\omega).$ 
The canonical quantization in these 
coordinates yields operators of the form
$$
\widehat{p_i} \, =\e \frac{\p}{\p q_i} ,\quad \widehat{q_i} = \frac1{\e} \, q_i
$$
on the Fock space of formal power series in $q_i$.
For any infinitesimal 
symplectic transformation~$A$
on $(\V,\omega)$ which satisfies
$$
\omega(A\,f,g)+\omega(f, A\,g) = 0 , \quad \forall \, f,g\in \V ,
$$
the Hamiltonian associated to~$A$ is defined by
$$H_{A}(f) = \frac12 \omega(f, A \, f ) = - \frac12 {\rm Res}_{z=\infty }  \,f(-z)\, A\,f(z)\, \frac{dz}{z^2} .$$
This Hamiltonian is a quadratic function on $\V$, and its quantization is defined via
$$ \widehat{p_i p_j} = \e^2\,\frac{\p^2}{\p q_i \p q_j} ,\quad \widehat{p_i q_j} = q_j \, \frac{\p}{\p q_i} , \quad \widehat{q_i q_j} = \frac1{\e^2} \, q_i q_j . $$
Denote the quantization of $H_{A}$ by~$\widehat{A}$. We have, for any two infinitesimal symplectic transformations 
$A,B$,  
$$\Bigl[\widehat{A} , \, \widehat{B}\Bigr] = \widehat{[A, B]}+\mathcal{C}\left(H_A , \, H_B\right) ,$$
where $\mathcal{C}$ is the so-called $2$-cocycle term satisfying
$$ 
\mathcal{C}(p_ip_j, q_k q_l) = -\mathcal{C}(q_k q_l, p_i p_j) = \delta_{i,k} \delta_{j,l}+\delta_{i,l} \delta_{j,k} , 
$$
and $\mathcal{C}=0$ for all other pairs of quadratic monomials in $p,q$.
 
Define the operators $l_k$ by
\beq
l_k = (-1)^{k+1} z^{3/2} \p_z^{k+1} z^{-1/2} ,\quad k\geq -1 .
\eeq
Then we have
\begin{lem} [\cite{Gi}] \label{g1}
The operators
$l_k$ are infinitesimal symplectic transformations on $\V$,  and their quantizations yield
the Virasoro operators defined in \eqref{virakdvintro1}--\eqref{virakdvintro3} as follows:
$$
L_k \left( \e^{-1}{\bf t}, \e \p/\p{\bf t}\right) = \left.\widehat{l_k}\right|_{q_i\mapsto  t_i, \,\p_{q_i} \mapsto \p_{t_i},\,i\geq 0} + \frac{\delta_{k,0}}{16} , \quad k\geq -1.
$$
\end{lem}

\begin{lem} [\cite{Gi}] \label{g2}
The multiplication operators $z^{1-2j},\,j\geq 1$ are infinitesimal symplectic transformations on $\V$, and the operators 
$D_j,\,j\geq 1$ defined in \eqref{dj} can be represented as
\beq\label{djgivental}
D_j = \left.\widehat{z^{1-2j}}\right|_{q_i\mapsto t_i, \, \p_{q_i} \mapsto \p_{t_i},\,i\geq 0}.
\eeq
\end{lem}

Consider now the quantization $\widehat\Phi$ of the symplectomorphism $f(z)\mapsto \Phi(z) f(z)$, 
where the function $\Phi(z)$ was defined by equation~\eqref{Gamma}. It will be defined by 
$$
\widehat{\Phi} =\left.e^{(\log \Phi(z))^{\wedge}}\right|_{q_i\mapsto t_i, \, \p_{q_i} \mapsto \p_{t_i},\,i\geq 0} ,
$$
where we replace $\log\Phi(z)$ by its asymptotic expansion as 
$|z|\to\infty$,  ${\rm Re}(z)\neq0$. The latter has the form, up to an inessential piecewise constant term
$$
\log \Phi(z) \,\sim\, \sum_{k=1}^\infty \frac{B_{2k}}{k(2k-1)} \frac{2^{-2k}-1}{z^{2k-1}} .
$$ 
Using \eqref{dj} we immediately arrive at the following lemma.
\begin{lem} \label{g3}
We have
\beq
\widehat{\Phi} = e^{\sum_{k=1}^\infty \frac{B_{2k}}{k(2k-1)}  \left(2^{-2k}-1\right) D_k}.
\eeq
\end{lem}

\begin{rmk}
The function $\Phi(z)$ is analytic near $z=0$, $\Phi(0)=1$ and
\beq\label{open}
\log \Phi(z) = -2z \log 2  - 2 \sum_{k=1}^\infty \frac{2^{2k}-1}{2k+1} \zeta(2k+1) \, z^{2k+1} , \quad |z|<\frac12 .
\eeq
One can define another quantum operator~$\hat\Phi_0$  by quantizing the series~\eqref{open}. Geometric interpretation of 
this quantum operator remains an interesting open question.
\end{rmk}
\end{appendices}

\noindent E-mails: \\
dubrovin@sissa.it\\
liusq@mail.tsinghua.edu.cn\\
diyang@ustc.edu.cn\\
youjin@mail.tsinghua.edu.cn

\end{document}